\documentclass[]{interact}
\usepackage{amsmath}
\usepackage{subcaption}
\usepackage{color}

\usepackage{ragged2e}
\theoremstyle{plain}

\theoremstyle{definition}

\theoremstyle{remark}

\newcommand{\phic}{\phi_\mathrm{c}}

\newcommand{\pD}[2]{\frac{\partial #1}{\partial #2}}

\newcommand{\phiu}{\phi_{u}}

\begin{document}
%
%
\title{Uncertainty Benchmarks for Time-Dependent Transport Problems}
%
%
\author{\name{W.~Bennett and R. G.~McClarren}
\affil{Dept.~Aerospace and Mechanical Engineering\\University of Notre Dame\\ 2006 St Joseph Dr, Notre Dame, IN 46556, USA}
\email{\{wbennet2,rmcclarr\}@nd.edu}
}
\maketitle

\begin{abstract}
       Verification solutions for uncertainty quantification are presented for time dependent transport problems where $c$, the scattering ratio,  is uncertain. The method of polynomial chaos expansions is employed for quick and accurate calculation of the quantities of interest and uncollided solutions are used to treat part of the uncertainty calculation analytically. We find that approximately six moments in the polynomial expansion are required to represent the solutions to these problems accurately. Additionally, the results show that if the uncertainty interval spans $c=1$, which means it is uncertain whether the system is multiplying or not, the confidence interval will grow in time. Finally, since the QoI is a strictly increasing function, the percentile values are known and can be used to verify the accuracy of the expansion. These results can be used to test UQ methods for time-dependent transport problems.
\end{abstract}

\section{Introduction} \label{sec:transport}\justifying
Uncertainty quantification (UQ) is necessary for any robust comparison between experimental data and accurate numerical simulations. In research fields concerning radiative flows, the accuracy of simulations has improved enough that  pragmatic researchers have begun to invest more effort into UQ \cite{smith2013uncertainty,mcclarren2018uncertainty}. The most intuitive choice with the least overhead for thus minded researchers is a Monte-Carlo sampling of whatever simulation method is on hand. The work of Fryer, et al. \cite{fryer2016uncertainties} is a good example of this so called non-intrusive method applied to laser-driven heat wave experiments in foams. It is also possible to bake uncertainty estimation processes into a simulation, like in \cite{kusch2020filtered}, which is an example of intrusive polynomial chaos expansions (PCE) that can be applied to time dependent radiation diffusion.

The bulk of research for UQ transport calculations published to date has come from the neutron transport community. This is a consequence of the interest nuclear engineers take in the effects of uncertainty on reactor criticality. Zheng and McClarren \cite{zheng2015variable} apply regression techniques to data sampled from a transport code to perform UQ on a simulated TRIGA reactor. The researchers in \cite{gilli2013uncertainty} apply PCE to a steady diffusion criticality problem. Williams \cite{williams2007polynomial} used a PCE treatment for a similar system, this time with the $P_1$ approximation.
Finally, Refs.~\cite{ayres2012uncertainty} and \cite{fichtl2011stochastic} apply PCE to true steady state transport calculations with uncertain cross sections. We  present a non-intrusive version of PCE that is an improvement on work previously surveyed in that it solves time dependent transport UQ problems.



For one dimensional, infinite medium, time dependent transport, the only physical parameters are the cross sections, the initial condition, and the source function. This study will not consider source uncertainty and wraps up uncertainties in the cross sections by defining the scattering ratio as an uncertain parameter. While PCE is often applied to the equations that govern a system, in this case the Green's function is known, although nontrivial to calculate, and given by Ganapol\cite{ganapol}; this solution was then extended to different sources in \cite{bennett2022benchmarks}. The organization of this work is as follows. Section \ref{sec:model} introduces the transport equation for the chosen configuration and corresponding benchmark solution and how uncertainty will be modeled. Section \ref{sec:PCE} discusses one dimensional PCE and how it has been applied to the problem and finally results are presented and discussed in Section \ref{sec:res}.



\section{Model Problem}\label{sec:model}
We begin with a problem of a planar pulse of neutrons in an infinite medium.
The time dependent, isotropic scattering transport equation in slab geometry with a Dirac delta function source in space and time is
\begin{equation}\label{eq:1dtransport}
    \left(\pD{}{t} + \mu\pD{}{x} + 1 \right)\psi(x,t,\mu) = \frac{c}{2}\,\phi(x,t) + \frac{1}{2}\, \delta(x)\delta(t),
\end{equation}
where  $\psi(x,t,\mu)$ is the angular flux and $\phi(x,t) = \int^1_{-1} \!d\mu'\,\psi(x,t,\mu')$ is the scalar flux; The spatial coordinate, $x \in \mathbb{R}$ is measured in mean-free paths from the origin, $\mu \in [-1,1]$ is the cosine of the angle between the polar angle of the direction of flight of a particle and the $x$-axis. $c$ is the average number of particles emitted (isotropically) from a particle collision. The scattering ratio, or the number of secondary particles emitted per collision, $c$, is defined as $c\equiv\sigma_s/\sigma_t$ where $\sigma_s$ is the sum of the scattering cross section and $\sigma_t$ is the total cross section. In the case where there is fission or $(\mathrm{n}, \nu\mathrm{n})$ reactions, the definition would change to include contributions from these reactions in the numerator.

The linearity of Eq.~\eqref{eq:1dtransport} allows the scalar flux to be divided into sum of the uncollided, $\phi_u$, and collided, $\phi_c$, parts ($\phi = \phi_u + \phi_c$) where the uncollided flux has not experienced scattering and the collided flux has.   gives The uncollided solution to Eq.~\eqref{eq:1dtransport} is \cite{ganapol}
\begin{equation}\label{eq:uncollided}
    \phi_u = \frac{1}{2}\frac{\exp\left(-t\right)}{t}\Theta\left(1-|\eta|\right),
\end{equation}
where $\Theta$ is the Heaviside step function. The collided solution also given in \cite{ganapol} is,
\begin{equation}\label{eq:pl_col}
    \phic(x,t) = c\left(\frac{e^{-t}}{8\pi}\left(1-\eta^2\right)\int^\pi_0\!du\, \mathrm{sec}^2\left(\frac{u}{2}\right)\mathrm{Re}\left[\xi^2e^{\frac{ct}{2}(1-\eta^2)\xi}\right]\right)\Theta(1-|\eta|),
\end{equation}
where
\begin{equation}\label{eq:xi}
    \xi(u,\eta) = \frac{\log q + i\, u}{\eta + i \,\mathrm{tan}(\frac{u}{2})},
\end{equation}
and
\begin{equation}\label{eq:q}
    q = \frac{1 + \eta}{1 - \eta}, \qquad \eta = \frac{x}{t}.
\end{equation}
Since the source in this configuration is a delta function, the only physical parameter is the scattering ratio, $c$. 

This solution is also extended in \cite{ganapol} into an axisymmetric cylindrical geometry to give the flux for a line source. In \cite{bennett2022benchmarks}, these solutions are used as integral kernels to produce solutions for other source configurations. Of those solutions, the square source and Gaussian source will be included in this study. The square source is a step source that is on for a time $t_0$ and is strength one inside of some width $x_0$ and zero outside. The Gaussian source is similar but the strength varies smoothly with some defined standard deviation, $\sigma$ (Not to be confused with the standard deviation of the solution or the random variable). 

This  introduces new physical parameters. For example, integrating Eqs.~\eqref{eq:uncollided} and \eqref{eq:pl_col} over a square source introduces the source width and source duration as parameters. While multi-dimensional chaos expansions could be used to include uncertainty in these parameters, this study will be restricted to uncertainty in $c$ only. 


To whit, $c$ is defined as an uncertain parameter, 
\begin{equation}
    c = \overline{c} + \omega_1 \theta,
\end{equation}
where $\overline{c}$ is a known mean value, $\omega_1$ is a constant, and $\theta$ is a uniform random variable, $\theta \sim \mathcal{U}[-1,1]$. This definition makes $\omega_1 >0$ give the width of the uncertain interval centered around the mean $\overline{c}$. It is noteworthy that if the uncertain interval extends from $c<1$ to $c>1$, there is uncertainty in whether the system supports long time behavior that is multiplying or decaying. Also, it should be noted that the uncollided flux (Eq.~\eqref{eq:uncollided}), has no uncertainty, a consequence of the uncollided particles being agnostic to the scattering properties of the material.

It is assumed that the magnitude of $\omega_1$ is small enough so that $c$ is always positive. For a uniform random variable, the probability density function (PDF) is defined as
\begin{equation}
f(\theta_i) = \begin{cases} \frac{1}{2} & \theta_i \in [-1,1] \\ 0 & \text{otherwise} \end{cases}. 
\end{equation}

With these choices for our uncertain parameters, the expectation of the scalar flux is defined as
\begin{equation}\label{eq:expectation_phi}
    E[\phi](x,t;c) = (E[\phi_u](x,t) + E[\phi_c](x,t;c)) = \phi_u(x,t) + \frac{1}{2}\int^1_{-1}\!d\theta_1 \, \phi_c(x,t;c). 
\end{equation}
The arguments are shown to indicate that the expected value of $\phi$ is a function of space and time. Additionally, the collided and total scalar flux are also parameterized by the value of $c$, as shown in Eq.~\eqref{eq:expectation_phi}. Henceforth, in the interest of notational parsimony, we will drop the $(x,t)$ arguments. 

The variance is also given by  
\begin{equation}
    \mathrm{VAR}[\phi] = E[\phi^2] - E[\phi]^2,
\end{equation}
which is in our case, 
\begin{equation}
    \mathrm{VAR}[\phi] = \frac{1}{2} \int_{-1}^1\!d\theta \:\phi_c(c)^2 -  \frac{1}{4}\left(\int_{-1}^1\!d\theta\:\phi_c(c)\right)^2.
\end{equation}
Higher moments (e.g. skewness, kurtosis, etc.) can also be calculated but will not be pursued in this study.


Percentile based statistics are also useful for describing a quantity of interest (QoI) in the presence of uncertainty and show greater resilience to outlier data than moment-based measures. Percentiles are calculated by finding the realization of the random variable that satisfies the relation
\begin{equation}
    p = \int_{-\infty}^x\!dx'\:f(x'),
\end{equation}
where $f$ is the PDF of the QoI and $p$ is the percentile represented in decimal form. These percentile values are calculated by estimating the cumulative distribution function (CDF) through sampling and then tabulating the inverse CDF. In this work, this is accomplished in Python with $\texttt{numpy}$'s $\texttt{quantile}$ function \cite{harris2020array}. 

\section{Polynomial Chaos}\label{sec:PCE}
The benchmark solutions referenced in the introduction are non-trivial to evaluate. For a Monte Carlo sampling of the QoI, the solution must be evaluated once for each realization of the uncertain parameter, which could require a large investment of computational time for an accurate solution. Direct integration could circumvent this difficulty in the calculation of moment based values, but the sometimes more useful percentile measures would not be available. 

If however, the solution is represented as a polynomial expansion in the uncertain variables and there are no discontinuities in the space of the random variable, evaluation will be relatively trivial once the coefficients have been calculated. This is one motivation for PCE. Therefore, the scalar flux as a function of the scattering ratio is approximated by, 
\begin{equation}
    \phi(c) = \phi_u + \sum_{j=0}^Na_j P_j(\theta), 
\end{equation}
where $P_j$ are Legendre polynomials.
The coefficients in the expansion are computed as
\begin{equation}\label{eq:expansion_coeffs}
    a_j = \frac{2j+1}{2}\int_{-1}^{1} d \theta \, \phi_c(c) P_j(\theta).
\end{equation}

These expansion coefficients can be efficiently calculated using widely-available quadrature routines. Once the expansion coefficients have been calculated, the orthogonality of the basis is invoked to write the moments of the expansion exactly. The expectation of the expansion is found to be exactly that of the function being approximated, in this case the scalar flux,
\begin{equation}\label{eq:expectation_PCE}
    E[\phi] = \phiu + a_0,
\end{equation}
since the definition of $a_0$ is identical to the expected value of the collided flux. 
Similarly, the variance simplifies to,
\begin{equation}\label{eq:var_exp}
    \mathrm{VAR}[\phi] =  \sum_{j=0}^N \frac{1}{2j +1}a_j^2 - a_0^2 = \sum_{j=1}^N \frac{1}{2j +1}a_j^2.
\end{equation}
The variance is not dependent on the uncollided solution, which is expected. Orthogonality also allows for higher moments of the expansion to be expressed exactly. 

Although the expansion coefficients give us a direct way to estimate moments of the distribution, they do not provide percentile information. To get this information we can sample values of the parameter $c$ from its underlying distribution and then evaluate the QoI using the polynomial representation. To efficiently calculate the percentiles  values of $c$ are sampled  via a quasi-random sequence, the Sobol sequence \cite{sobol1967distribution}, taking advantage of the faster convergence possible with these sequences. Roughly one million samples were taken for each problem. The code to reproduce the results shown below is freely available\footnote{github.com/wbennett39/pce$\_$transport}.

\section{Results}\label{sec:res}
\begin{figure}
    \centering
    \includegraphics{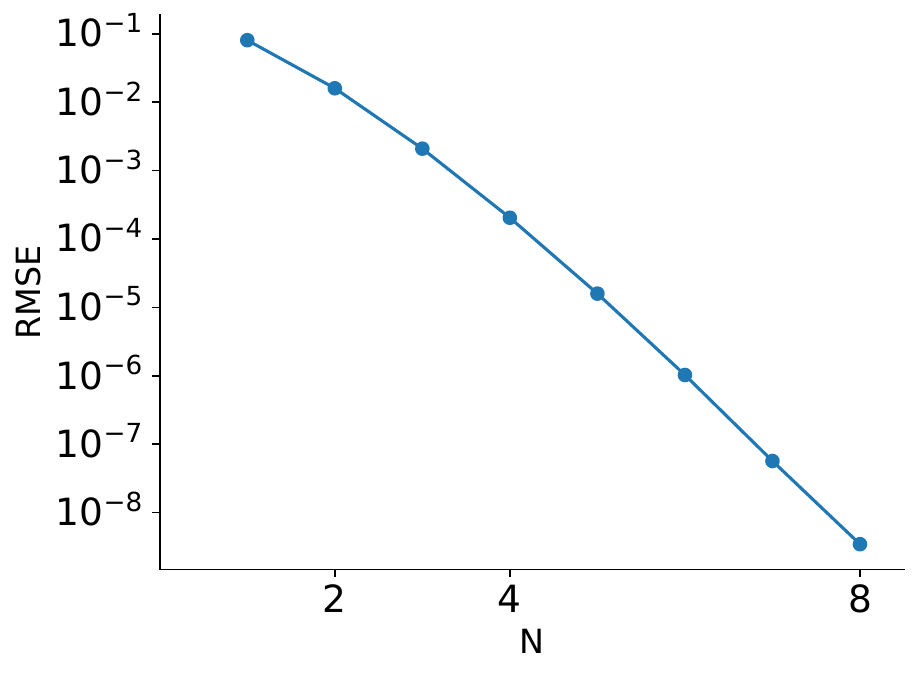}
    \caption{Log linear scaled root mean square error (RMSE) of the expansion variance versus expansion order N, calculated with Eq.~\eqref{eq:var_exp} for $t=5$, $c=1$, $a_1=0.5$.}
    \label{fig:converge}
\end{figure}

For these results to have any value, the polynomial chaos expansion must be a good approximation of the true function. Figure \ref{fig:converge} provides a convergence test of the variance of the scalar flux, calculated with Eq.~\eqref{eq:var_exp} against an analytic expression for the variance for the plane pulse problem. The constant slope on the log-linear plot test indicates geometric convergence. Since the first moment is exactly the analytic expression, we can be confident that an order $\approx 6$ basis is a good representation of the true solution for this problem since it represents the first two moments well when the uncertainty in the scattering ratio is large (fifty percent).

For four problems, three in slab geometry and one in cylindrical, plots are given at early and intermediate times of moment and percentile based statistical measures for a range of $\overline{c}$. Table \ref{tab:1} gives the problem setup for each figure. Each result shows a similar trend in the $\overline{c}=1$ cases of a widening of the certainty interval as time progresses. This is due to the difference between the exponential decay behavior of a non-multiplying system and the exponential growth of a multiplying system. 

\begin{figure}
    \centering
    \begin{subfigure}[b]{0.48\textwidth}
    \includegraphics[width=\textwidth]{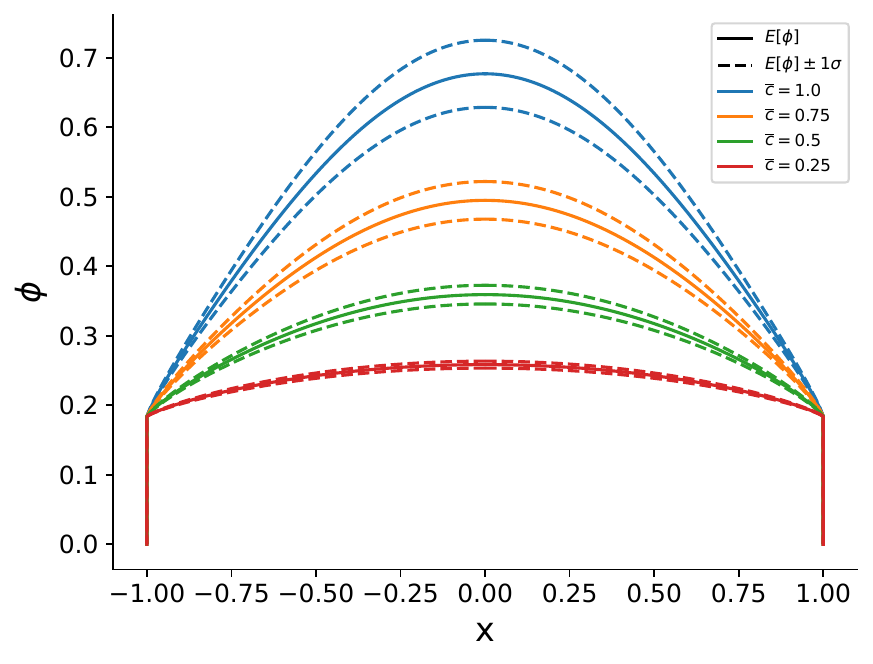}
    \caption{$t=1$ moments}
    \end{subfigure}
    \begin{subfigure}[b]{0.48\textwidth}
    \includegraphics[width=\textwidth]{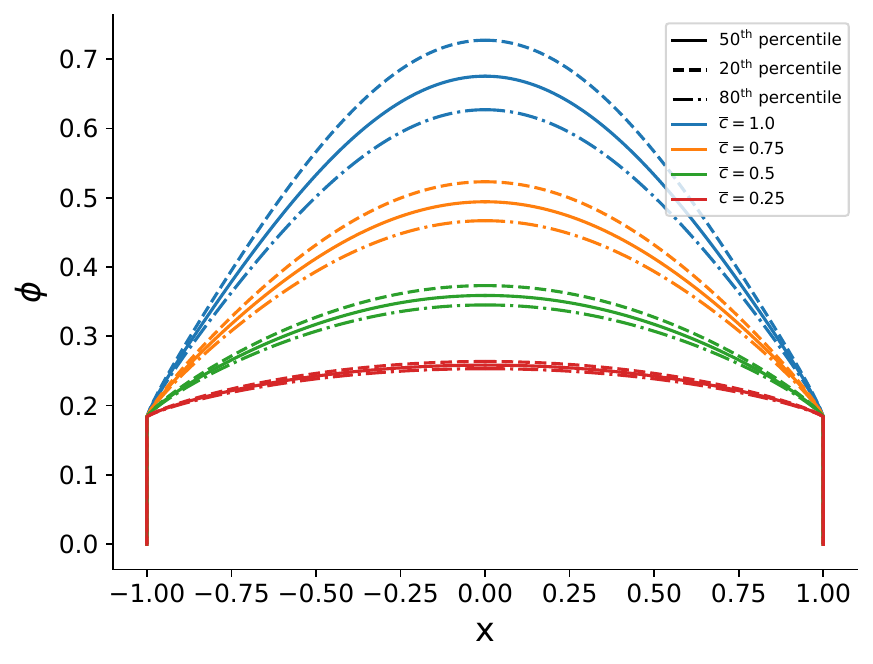}
    \caption{$t=1$ quantiles}
    \end{subfigure}
    \begin{subfigure}[b]{0.48\textwidth}
    \includegraphics[width=\textwidth]{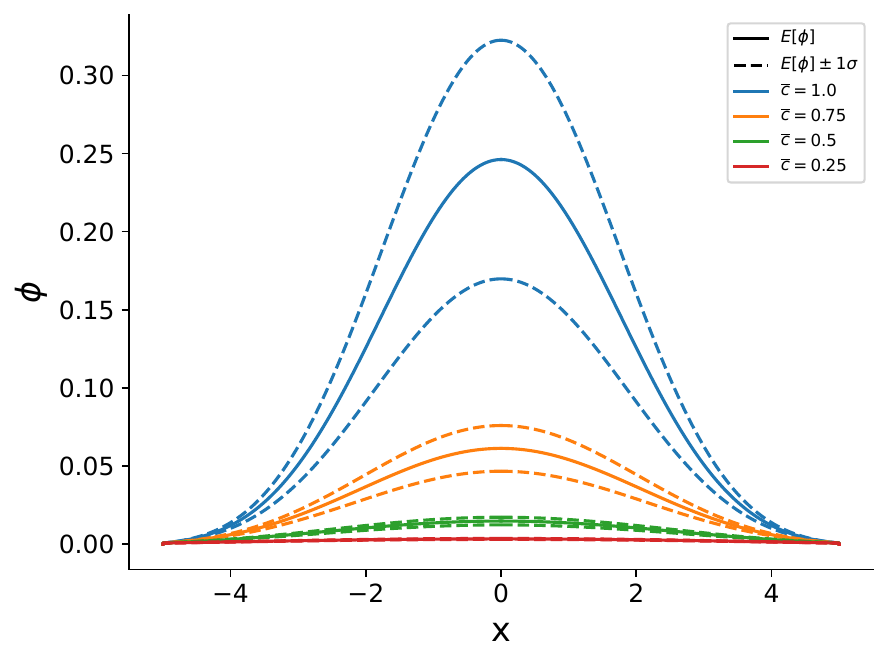}
    \caption{$t=5$ moments}
    \end{subfigure}
    \begin{subfigure}[b]{0.48\textwidth}
    \includegraphics[width=\textwidth]{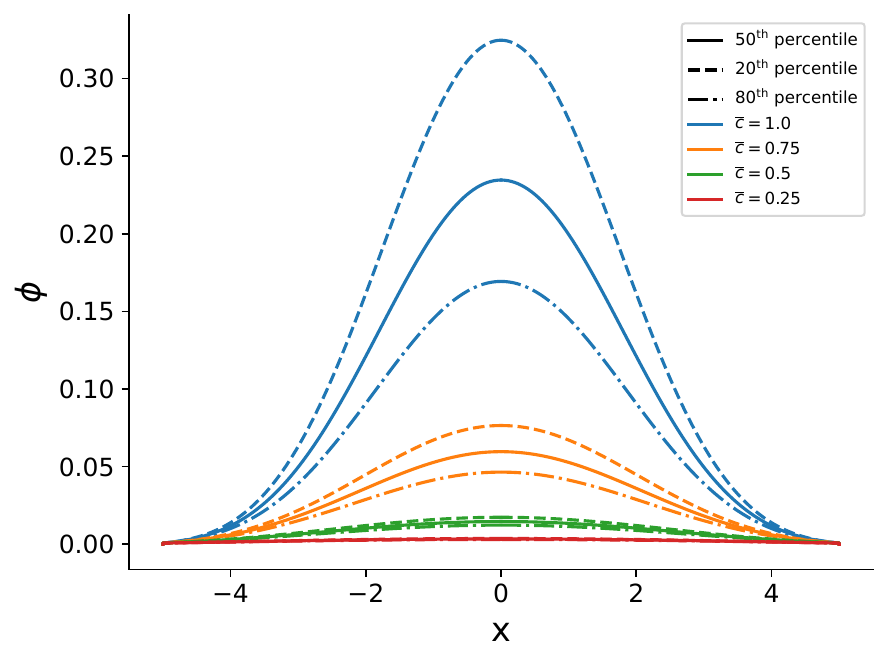}
    \caption{$t=5$ quantiles}
    \end{subfigure}
   \caption{Moment (left) and quantile (right) descriptions from a $N = 6$ order expansion of the plane pulse transport solution with a 10\% uncertainty  ($\omega=0.1\overline{c}$) in $c$ at $t=1$ (top) and $t=5$ (bottom).}
    \label{fig:t1_moments}
\end{figure}

\begin{figure}
    \centering
    \begin{subfigure}[b]{0.48\textwidth}
    \includegraphics[width=\textwidth]{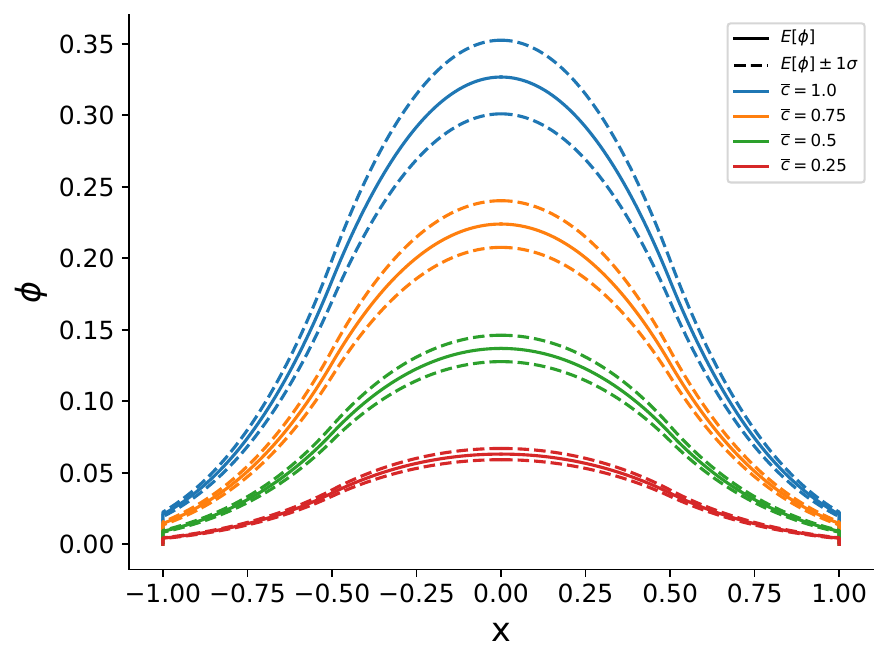}
    \caption{$t=1$ moments}
    \end{subfigure}
    \begin{subfigure}[b]{0.48\textwidth}
    \includegraphics[width=\textwidth]{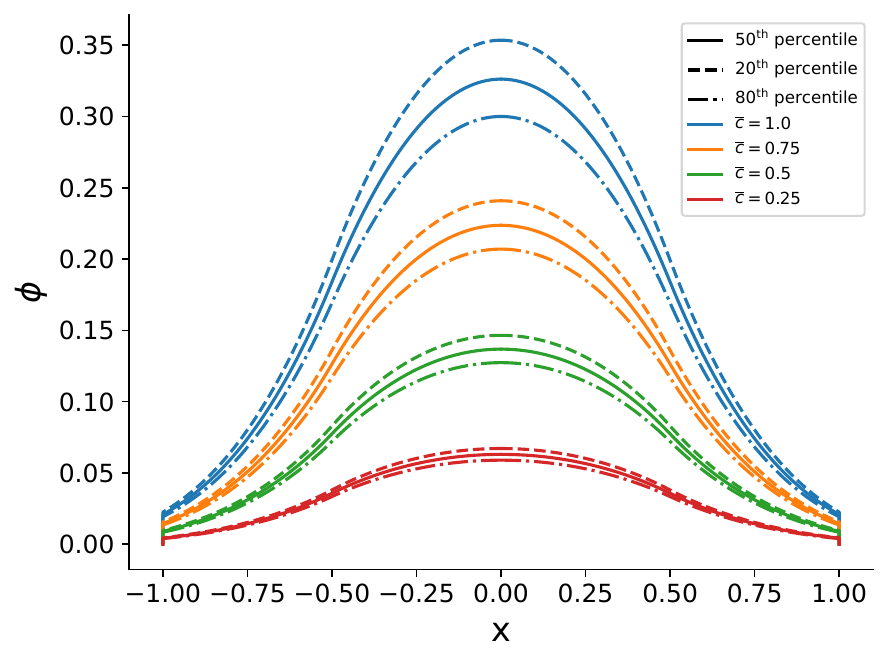}
    \caption{$t=1$ quantiles}
    \end{subfigure}
    \begin{subfigure}[b]{0.48\textwidth}
    \includegraphics[width=\textwidth]{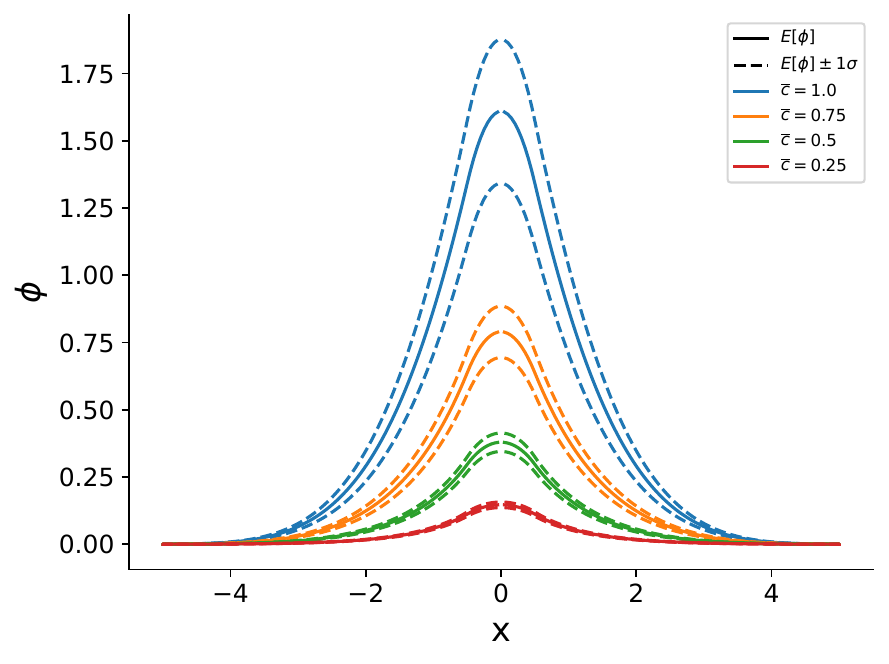}
    \caption{$t=5$ moments}
    \end{subfigure}
    \begin{subfigure}[b]{0.48\textwidth}
    \includegraphics[width=\textwidth]{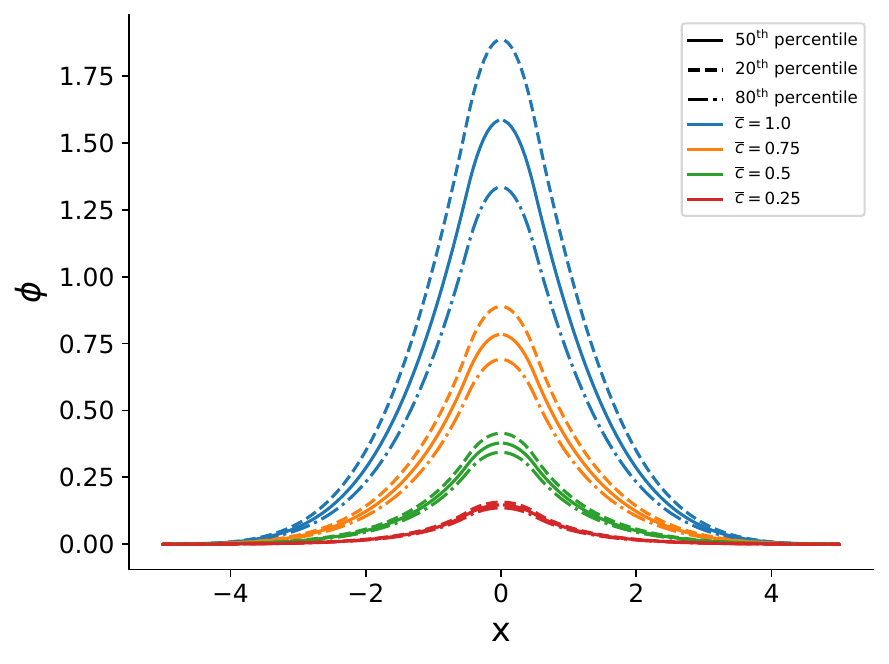}
    \caption{$t=5$ quantiles}
    \end{subfigure}
   \caption{Moment (left) and quantile (right) descriptions from a $N = 6$ order expansion of the square source transport solution with a 10\% uncertainty  ($\omega=0.1\overline{c}$) in $c$ at $t=1$ (top) and $t=5$ (bottom).}
    \label{fig:t1_moments_square}
\end{figure}

\begin{figure}
    \centering
    \begin{subfigure}[b]{0.48\textwidth}
    \includegraphics[width=\textwidth]{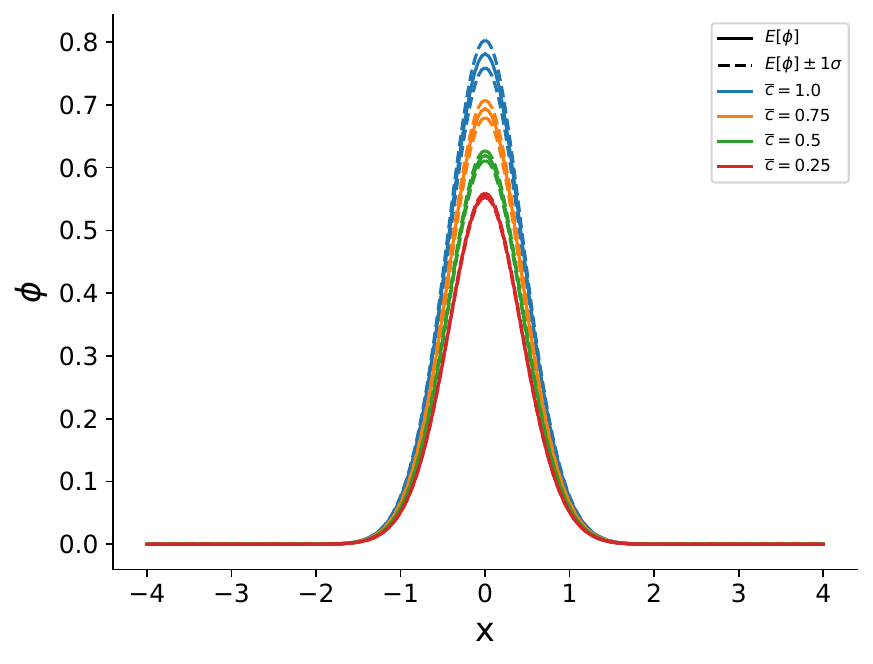}
    \caption{$t=1$ moments}
    \end{subfigure}
    \begin{subfigure}[b]{0.48\textwidth}
    \includegraphics[width=\textwidth]{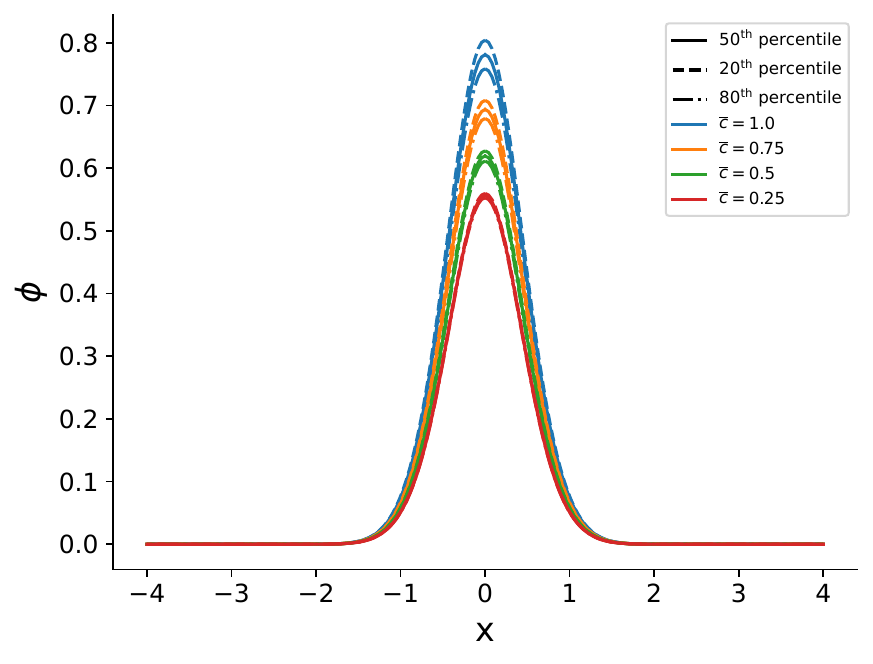}
    \caption{$t=1$ quantiles}
    \end{subfigure}
    \begin{subfigure}[b]{0.48\textwidth}
    \includegraphics[width=\textwidth]{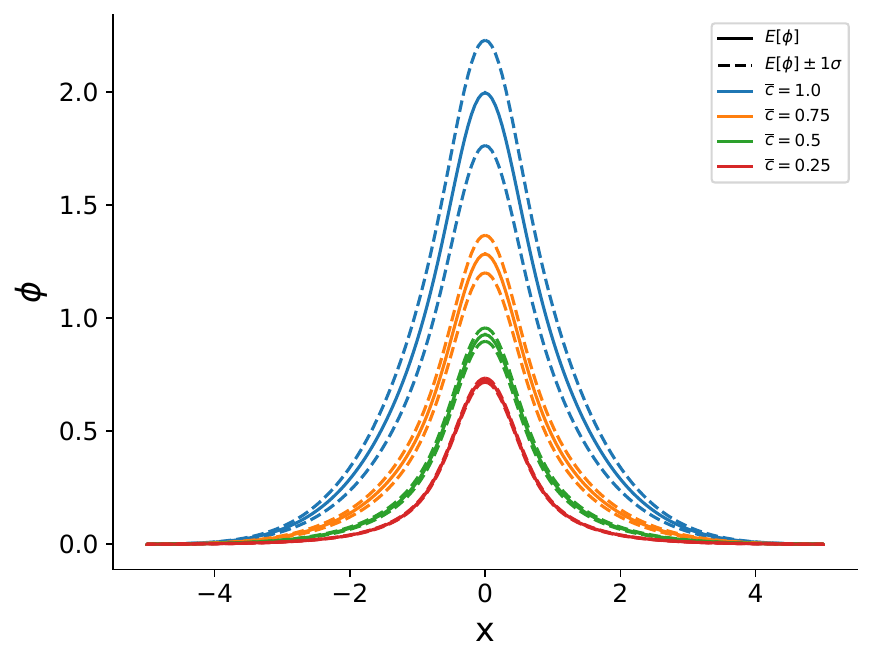}
    \caption{$t=5$ moments}
    \end{subfigure}
    \begin{subfigure}[b]{0.48\textwidth}
    \includegraphics[width=\textwidth]{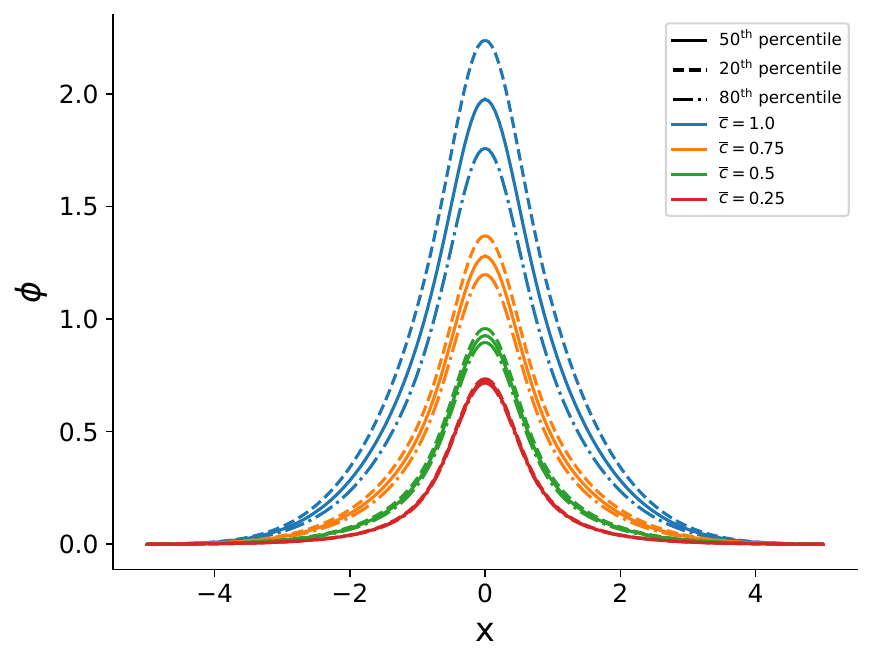}
    \caption{$t=5$ quantiles}
    \end{subfigure}
   \caption{Moment (left) and quantile (right) descriptions from a $N = 6$ order expansion of the Gaussian source transport solution with a ten percent uncertainty  ($\omega=0.1\overline{c}$) in $c$ at $t=1$ (top) and $t=5$ (bottom).}
    \label{fig:t1_moments_gaussian}
\end{figure}

\begin{figure}
    \centering
    \begin{subfigure}[b]{0.48\textwidth}
    \includegraphics[width=\textwidth]{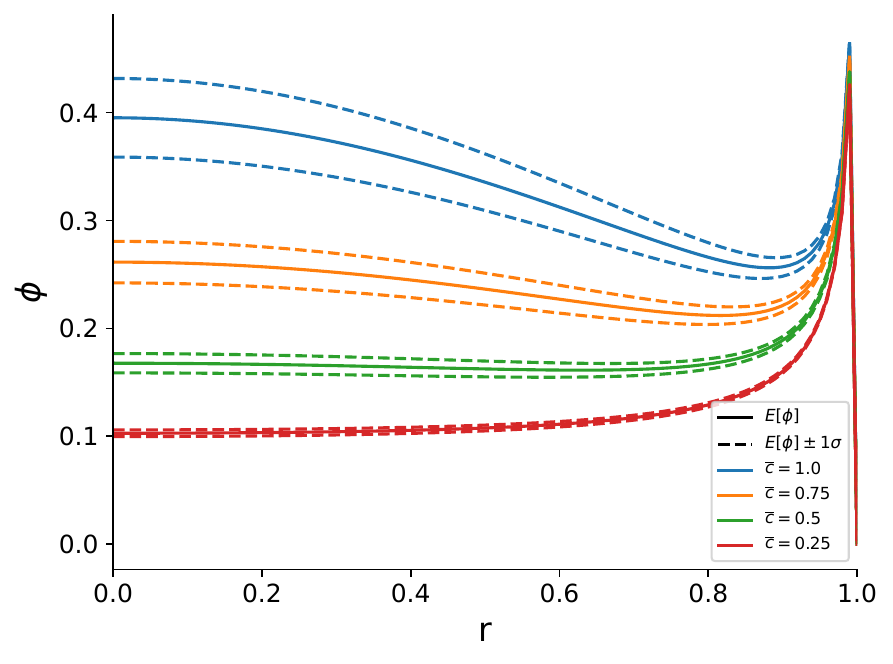}
    \caption{$t=1$ moments}
    \end{subfigure}
    \begin{subfigure}[b]{0.48\textwidth}
    \includegraphics[width=\textwidth]{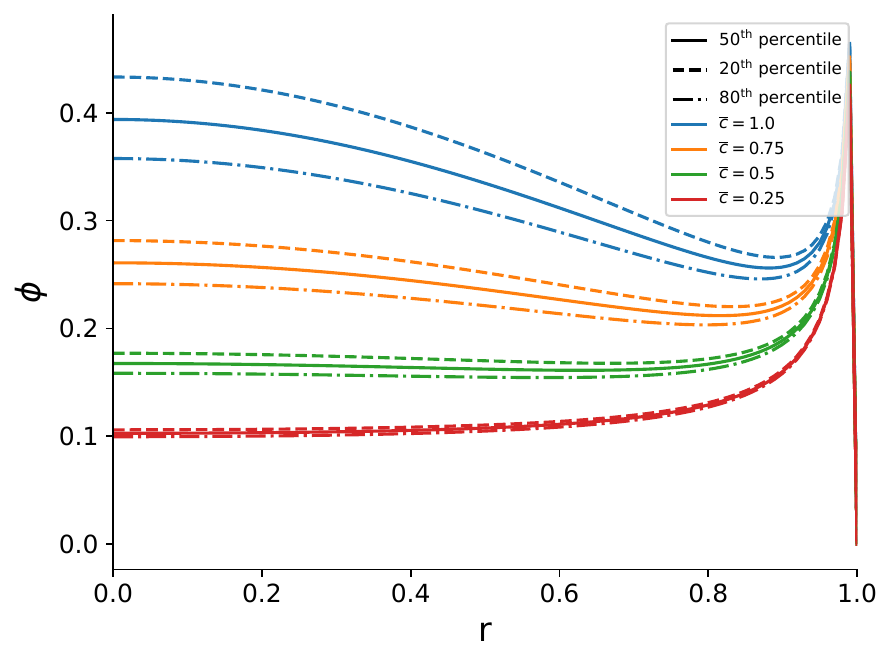}
    \caption{$t=1$ quantiles}
    \end{subfigure}
    \begin{subfigure}[b]{0.48\textwidth}
    \includegraphics[width=\textwidth]{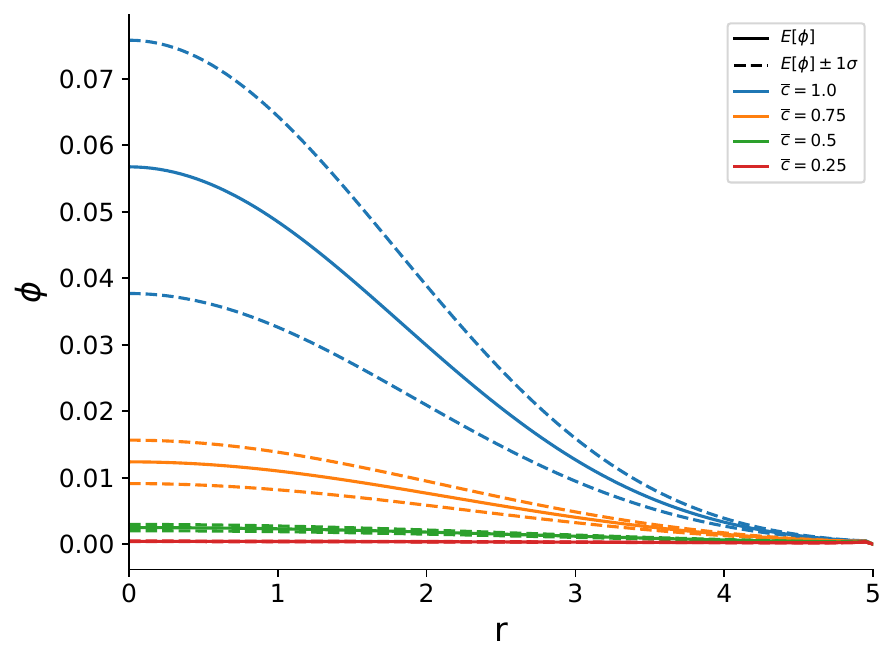}
    \caption{$t=5$ moments}
    \end{subfigure}
    \begin{subfigure}[b]{0.48\textwidth}
    \includegraphics[width=\textwidth]{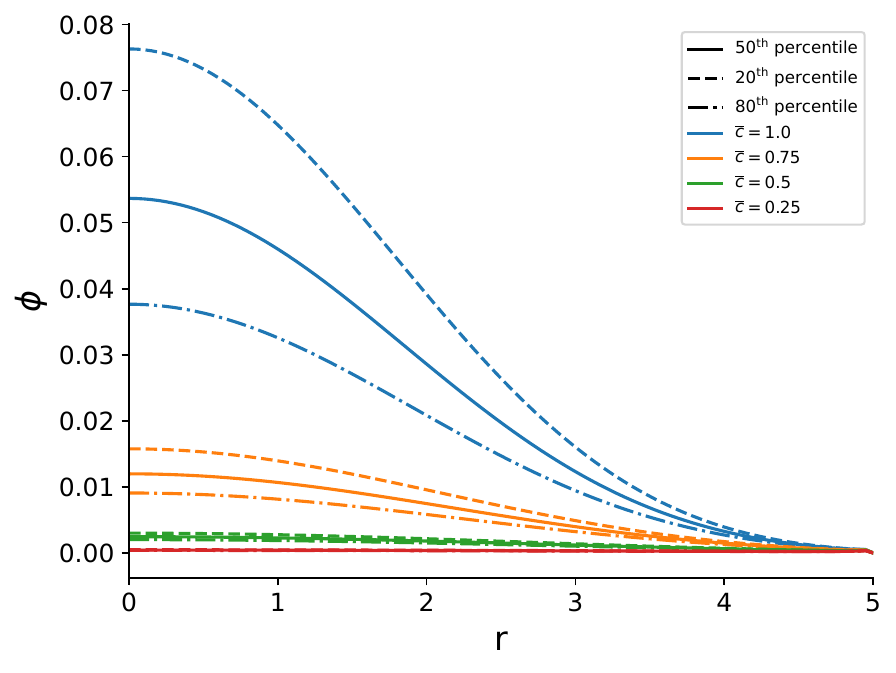}
    \caption{$t=5$ quantiles}
    \end{subfigure}
   \caption{Moment (left) and quantile (right) descriptions from a $N = 6$ order expansion of the line source transport solution with a 10\% uncertainty  ($\omega=0.1\overline{c}$) in $c$ at $t=1$ (top) and $t=5$ (bottom).}
    \label{fig:t1_moments_line}
\end{figure}

\begin{figure}
    \centering
    \includegraphics{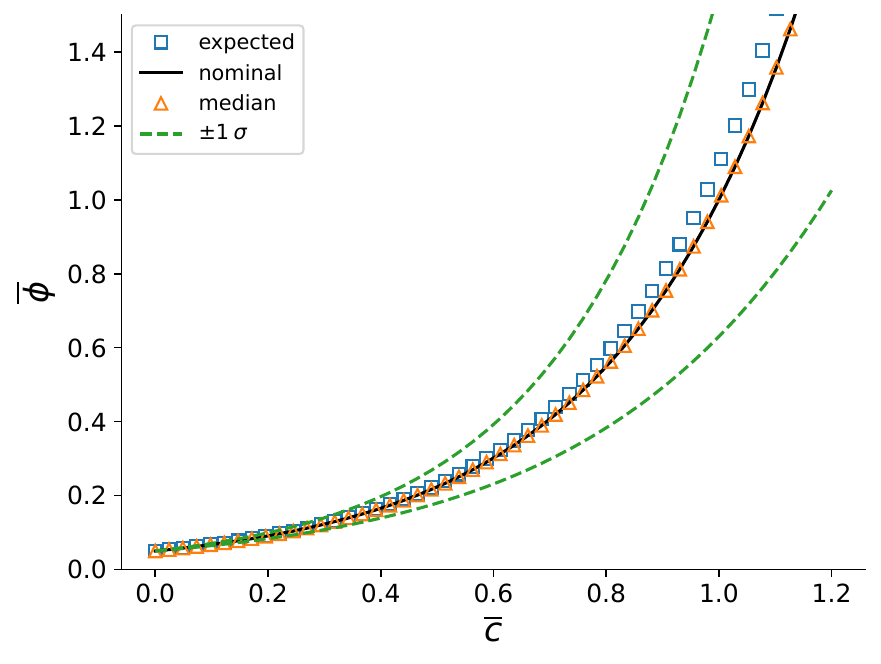}
    \caption{Total mass of the system for the plane pulse problem at three mean free times with 25\% uncertainty vs mean scattering ratio, $\overline{c}$ for the nominal case ($c=\overline{c}$), given by Eq.~\eqref{eq:mass_analytic}, the expected value and $\pm 1\sigma$ given by a $N=6$ expansion. The median is calculated with $\approx$ $10^6$ samples.}
    \label{fig:energy_vs_c}
\end{figure}

Compared to the other problems, the Gaussian source, Figure \ref{fig:t1_moments_gaussian} shows relatively smaller ranges where the solution could safely assumed to be between. This could be because the solution is smooth at early times. The other figures (Figures \ref{fig:t1_moments}, \ref{fig:t1_moments_square}, \ref{fig:t1_moments_line}) show structure at early times and then relax into a Gaussian shape at later times. For each problem, the expected value seems to be the same as the median. They are slightly different however and diverge as time progresses. Figure \ref{fig:energy_vs_c} shows this phenomenon. While the uncertainty goes to zero at the wavefront in Figure \ref{fig:t1_moments}, it will not in Figure \ref{fig:t1_moments_square} since the collided solution at the wavefront is nonzero. All results also follow the trend of having the highest variance at center of the problem ($x=0$ for slab problems and $r=0$ for cylindrical).

The total ``mass'' of the system, which is the integral over all space of the solution,
\begin{equation}\label{eq:mass_integral}
    \overline{\phi}(t;c) = \int_{-\infty}^\infty\!dx'\phi(c,x',t)
\end{equation}
can yield important insights for these types of problems. The solution to Eq.~\eqref{eq:mass_integral} for the plane pulse is,
\begin{equation}\label{eq:mass_analytic}
    \overline{\phi}(t;c) = \exp\left(t(c-1)\right).
\end{equation}
This is the mass of the system with no uncertainty. The masses of some statistical descriptions of the system (expected, median) are shown at $t=3$ for a twenty five percent uncertainty in Figure \ref{fig:energy_vs_c} for the plane pulse problem. Three mean free times was chosen as an intermediate value between early times when the expected value shows little deviation from the nominal value ($c=\overline{c}$) and later times when they are highly disparate. 

\begin{figure}
    \centering
    \includegraphics{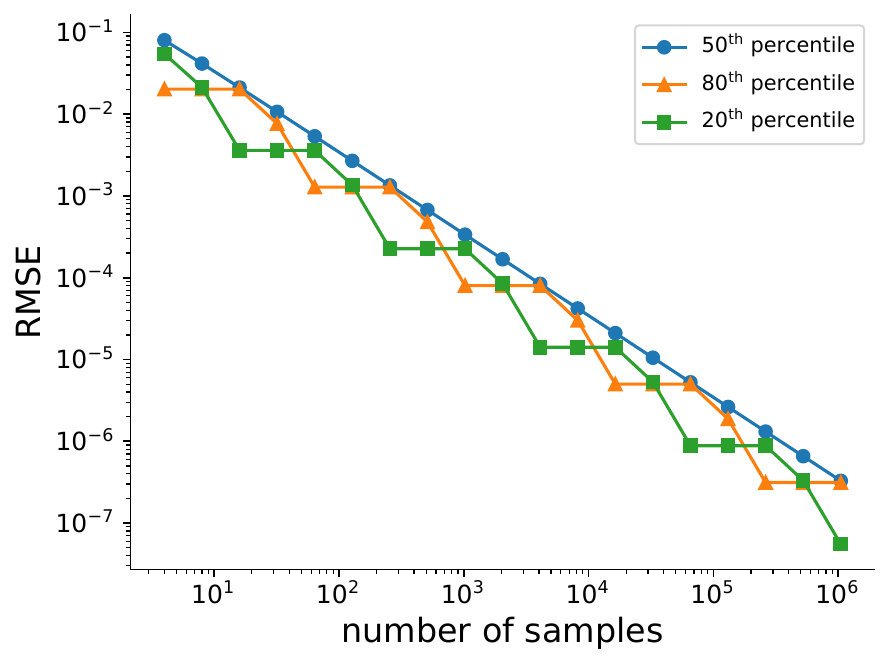}
    \caption{The RMSE for the plane-pulse case between a sampled $N=8$ expansion and the exact solution evaluated at the $p^\mathrm{th}$ percentile of the random variable. In this case $\overline{c}=1.1$, there is a 25\% uncertainty, and the evaluation time is $t=5$. }
    \label{fig:samp_converge}
\end{figure}

On the scale of the plot (Figure \ref{fig:energy_vs_c}), the median and nominal ($c=\overline{c}$) curves for the system mass appear coincident. There is a reason for this. For a strictly increasing function of a random variable, it can be shown that a quantile realization of function, the $p^\mathrm{th}$ percentile, is the same as the function evaluated at the $p^\mathrm{th}$ percentile of the random variable. This means that in this case, the $50^\mathrm{th}$ percentile of the scalar flux is equal to the case of no uncertainty, the nominal value. A derivation of this is included in Appendix \ref{app:proof}. All this is to say, sampling the expansion to estimate quantile values is not actually necessary in this case. However, it is a good test of the method to check if the sampled expansion percentile values converge to the known values, which we show in Figure \ref{fig:samp_converge}. The order of the expansion is increased to eight for this plot to show convergence past the $N=6$ floor at $\approx10^{-6}$. We do not expect spectral convergence in this case, since the random sampling method at best converges algebraically. 


\begin{table}[]
\caption{Description of uncertainty benchmarks, each with $c=\bar{c} + \omega_1 \theta$ as the uncertain parameter}
{ \begin{tabular}{lllllllllllll} 
Source                               & functional form                                       & parameters                &  figure number(s) &  &  &  &  &  &  &  &  \\ \cline{1-5}
\multicolumn{1}{l|}{plane pulse}     & $\delta(x)\delta(t)$                                  & -                         &  \ref{fig:converge}, \ref{fig:t1_moments}  \ref{fig:energy_vs_c}, \ref{fig:samp_converge}        &  &  &  &  &  &  &  &  \\
\multicolumn{1}{l|}{square source}   & $\Theta(x_0-|x|)\Theta(t_0-t)$                        & $x_0 = 0.5, \: t_0 = 5$   & \ref{fig:t1_moments_square}             &  &  &  &  &  &  &  &  \\
\multicolumn{1}{l|}{Gaussian source} & $\exp\left(\frac{-x^2}{\sigma^2}\right)\Theta(t_0-t)$ & $\sigma= 0.5, \: t_0 = 5$ & \ref{fig:t1_moments_gaussian}             &  &  &  &  &  &  &  &  \\
\multicolumn{1}{l|}{line source}     & $\delta(r)\delta(t)$                                  & -                         & \ref{fig:t1_moments_line}             &  &  &  &  &  &  &  &  \\
                                     &                                                       &                           &                                     &               &  &  &  &  &  &  &  &  \\
                                     &                                                       &                           &                                     &               &  &  &  &  &  &  &  & 
\end{tabular}
}
\label{tab:1}
\end{table}




\section{Conclusions}
The method of polynomial chaos expansions has been applied to infinite medium isotropic scattering benchmark problems with uniform uncertainty in the scatting ratio to produce verification solutions for transport uncertainty quantification methods and codes. While simulation codes are usually more easily adaptable to finite medium problems, the finite information speed of transport allows any infinite medium problem to be treated like a finite medium problem that has not yet interacted with the boundary.

While one dimensional uncertainty benchmarks do not test the ability of a method to calculate the aggregate effects of uncertainty in multiple parameters, the simple problems given here are a good starting point and are certainly nontrivial, especially when the uncertainty spans  $\overline{c}=1$. It is also shown that since the scalar flux is a strictly increasing function of $c$, the sampled percentile values of the scalar flux will converge with increasing sample size to the scalar flux evaluated at the desired percentile values of the random variable.
 
In terms of future work, this method can be extended into higher dimensions where the source width, duration, etc. are uncertain. In addition to creating benchmarks for uncertain transport problems, PCE can be employed to produce benchmark solutions for a variety of problems, such as multiplying media with delayed neutrons, or radiative transfer problems such as the Su-Olson problem \cite{mcclarren2008analytic}, the Marshak wave \cite{marshak,petschek1960penetration}, or any transport/ radiative transfer problem that has a benchmark solution. 

\section*{ACKNOWLEDGEMENTS}

This work was supported by a Los Alamos National Laboratory
under contract \#599008, “Verification for Radiation Hydrodynamics Simulations”.


\bibliographystyle{unsrt}
\bibliography{ref}

\appendix
\section{Analytic expectation for plane pulse}\label{app:a1}
For the plane pulse problem, the expectation of the collided flux is given by
\begin{multline}
    E[\phi_c] = \frac{1}{2}\int_{-1}^1\!d\theta\:\phi_c(\theta)=\frac{1}{8 \pi  a_1 \left(\eta ^2-1\right) t^2} \int_0^{\pi}\!du\: \sec ^2\left(\frac{u}{2}\right)\mathrm{Re}\bigg{[}  e^{-\frac{1}{2} \left(\eta
   ^2-1\right) \xi  t \left(a_1+\overline{c}\right)-t}\times 
   \\ \left(\left(\eta ^2-1\right) \xi  t
   \left(a_1+\overline{c}\right)+e^{a_1 \left(\eta ^2-1\right) \xi  t} \left(\left(\eta ^2-1\right)
   \xi  t \left(a_1-\overline{c}\right)-2\right)+2\right)\bigg{]}\Theta (1-| \eta | ).
\end{multline}
\section{Strictly increasing function of a random variable}\label{app:proof}
Let $\mathbb{F}_g(w)$ be the CDF for the uncertain function $g$ with a single random variable input evaluated at $w$.  $\mathbb{F}_\phi$ can be related to the CDF of the random variable, $\mathbb{F}_\theta$,
\begin{equation}\label{eq:CDF relation}
    \mathbb{F}_\phi(w) = \mathbb{F}_\theta\left(\phi^{-1}(w)\right).
\end{equation}
This relation can be used to show that the $p$th percentile of $\phi$ is the same as evaluating $\phi$ at the $p$th percentile of the random variable. First, define two $w$'s so that, 
\begin{equation}
    \mathbb{F}_\phi(w_1) = p \qquad \mathbb{F}_\theta(w_2) = p.
\end{equation}
That is, each $w$ returns $p$ when it is an input in to the respective CDF. Therefore,
\begin{equation}
     \mathbb{F}_\phi(w_1) = \mathbb{F}_\theta(w_2).
\end{equation}
Applying Eq.~\eqref{eq:CDF relation},
\begin{equation}
     \mathbb{F}_\theta(\phi^{-1}(w_1)) = \mathbb{F}_\theta(w_2).
\end{equation}
Since the CDF must be one-to-one if $\phi(c)$ is a bijective function in $c$, we have
\begin{equation}
    \phi^{-1}(w_1) = w_2.
\end{equation}
After inverting, we find,
\begin{equation}\label{eq:end_proof}
    w_1 = \phi(w_2).
\end{equation}
We have shown that the $p$th percentile of the function $\phi$, $w_1$, is equal to evaluating $\phi$ at $w_2$, which is the $p$th percentile of the random variable.


\end{document}